\documentclass[twocolumn,showpacs,preprintnumbers,amsmath,amssymb]{revtex4}

\usepackage{graphicx}
\usepackage{dcolumn}
\usepackage{bm}

\begin{document}

\preprint{APS/123-QED}

\title{Effect of Elastic Deformations on the Critical Behavior
of Disordered Systems with Long-Range Interactions}

\author{S.V. Belim}
 \email{belim@univer.omsk.su}
\affiliation{%
Omsk State University, 55-a, pr. Mira, Omsk, Russia, 644077
\textbackslash\textbackslash
}%

\date{\today}

\begin{abstract}
A field-theoretic approach is applied to describe behavior of
three-dimensional, weakly disordered, elastically isotropic,
compressible systems with long-range interactions at various
values of a long-range interaction parameter.
Renormalization-group equations are analyzed in the two-loop
approximation by using the Padé–Borel summation technique. The
fixed points corresponding to critical and tricritical behavior of
the systems are determined. Elastic deformations are shown to
changes in critical and tricritical behavior of disordered
compressible systems with long-range interactions. The critical
exponents characterizing a system in the critical and tricritical
regions are determined.

\end{abstract}

\maketitle

The effect of a long-range interaction described by a
power law $r^{– D – \sigma}$ was studied analytically in the framework of
the $\varepsilon$-expansion [1–3] and numerically by Monte Carlo methods [4–6]
for two- and one-dimensional systems. It was shown that effects
due to long-range interaction are essential for the critical
behavior of Ising systems when $\sigma < 2$. The two-loop approximation
applied in [7] directly in the three-dimensional space
corroborates the predictions of the $\varepsilon$-expansion for homogeneous
systems with long-range interactions.

In structural phase
transitions without piezoelectric effect in the paraphase, elastic
deformations play the role of a secondary order parameter whose
fluctuations are not critical in most cases [8, 9]. Since the main
contribution to striction effects in the critical region is due to
the distance dependence of the exchange integral, only isotropic
elastic systems are discussed below.

It was shown in [10, 11] that
coupling between order-parameter fluctuations and elastic
deformations can be responsible both for change in critical
behavior and for emergence of tricritical and tetracritical
points. The introduction of frozen point impurities into a system
both changes the critical behavior and eliminates multicritical
points [12]. The analysis presented in [13] showed that frozen
structural defects introduced into spin systems with long-range
interactions increased the value of a long-range parameter
corresponding to the transition to mean-field critical behavior.
The effect of elastic deformations on homogeneous systems with
long-range interactions also changes the critical behavior [14].
Therefore, it would be interesting to analyze the combined effect
of structural defects and elastic deformations on systems with
long-range interactions.

In this work, the critical behavior of disordered
compressible systems with long-range interactions in
the three-dimensional space is described for various
values of the long-range parameter $\sigma$.

The Hamiltonian of the disordered Ising model
including elastic deformations and long-range interactions
can be written as
\begin{eqnarray}\label{gam1}
&&H_0=\frac 12\int d^Dq(\tau _0+q^\sigma)S_qS_{-q}
+\frac 12\int d^Dq\Delta \tau _qS_qS_{-q}\nonumber \\
&&+u_0\int d^DqS_{q1}S_{q2}S_{q3}S_{-q1-q2-q3}\nonumber\\
&&+a_3\int d^Dqy_{q1}S_{q2}S_{-q1-q2} \\
&&+\frac{a_3^{(0)}}\Omega y_0\int d^DqS_qS_{-q}
+\frac 12a_1 \int d^Dqy_qy_{-q}\nonumber\\
&&+\frac 12\frac{a_1^{(0)}}\Omega y_0^2+\int
d^Dqh_qy_q+\frac{h_0}\Omega y_0  \nonumber
\end{eqnarray}
where $S_q$ is an order parameter, $u_0$ is a positive constant,
$\tau_0\sim|T-T_c|/T_c$, $T_c$ is the phase-transition temperature,
$\sigma$ is the long-range parameter, $\Delta\tau_q$ is a random impurity
field (e.g., random temperature), $a_1$ and $a_2$ are the elastic
constants of a crystal, and $a_3$ is the quadratic striction
constant. The coupling between impurities and the
nonfluctuating order parameter $y(x)=\sum\limits_{\alpha =1}^3u_{\alpha\alpha}(x)$,
where $u_{\alpha \beta}$ is the strain tensor, is specified by the random
field $h_q$ thermodynamically conjugate to $u_{\alpha\alpha }(x)$
In Eq. (1), integration is performed in the terms depending
on nonfluctuating variables that are not coupled to the
$S_q$, and the $y_0$ terms (describing homogeneous deformations) are separated.
It was shown in [8] that this separation is necessary because inhomogeneous
deformations $y_q$ are responsible for exchange of acoustic phonons and for
long-range effects that are absent under homogeneous deformations.
When the impurity concentration is low, random fields $\Delta\tau_q$, $h_q$ and $h_0$
can be treated as Gaussian and specified by the function
\begin{eqnarray}
&&P[\Delta \tau ,h,h_0]=A\exp [ -\frac 1{8b_1}\int \Delta \tau_q^2d^Dq\nonumber\\
&&-\frac 1{8b_2}\int h_q^2d^Dq
-\frac 1{8b_3}\int h_0d^Dq\\
&&-\frac1{4b_4}\int \Delta \tau _qh_qd^Dq -\frac 1{4b_5}\int
\Delta \tau_qh_0d^Dq],\nonumber
\end{eqnarray}
where $A$ is a normalization factor and $b_i$ denotes positive
constants proportional to the concentration of frozen
structural defects.

Using the replica procedure to average over the random
fields generated by frozen structural defects, we
obtain an effective Hamiltonian of the system:
\begin{eqnarray}
&&H_{R}=\frac 12\int d^Dq(\tau _0+q^\sigma)\sum\limits_{a=1}^mS^a_qS^a_{-q}\nonumber\\
&&-\frac{\delta _0}2\sum\limits_{a,b=1}^m\int d^Dq(
S^a_{q1}S^a_{q2})(S^b_{q3}S^b_{-q1-q2-q3}) \nonumber \\
&&+u_0\sum\limits_{a=1}^m\int d^DqS^a_{q1}
S^a_{q2}S^a_{q3}S^a_{-q1-q2-q3}\nonumber \\
&&+g_0\sum\limits_{a=1}^m\int d^Dqy^a_{q1}S^a_{q2}
S^a_{-q1-q2}\\
&&+\frac{g_0^{(0)}}\Omega \sum\limits_{a=1}^my^a_0\int
d^DqS^a_qS^a_{-q} +\frac 12\lambda \int d^Dqy_qy_{-q}+\frac
12\frac{\lambda _0}\Omega y_0^2\nonumber
\end{eqnarray}
where the positive constants $\delta _0$, $g_0$,
$g_0^{(0)}$, $\lambda$ and $\lambda _0$  can
be expressed in terms of $a_i$ and $b_i$. The properties of the
original system can be obtained in the limit $m\rightarrow0$,
where m is the number of replicas (images).

Define an effective Hamiltonian depending only on
the strongly fluctuating order parameter $S_q$ by the relation
\begin{eqnarray}\label{usr}
\exp \{-H[S]\}=B\int \exp \{-H_{R}[S,y]\}\prod dy_q
\end{eqnarray}
If an experiment is carried out at constant volume, then
$y_0$ is a constant and integration in Eq. (4) is performed
over inhomogeneous deformations, whereas homogeneous
deformations do not contribute to the effective
Hamiltonian. At constant pressure, the Hamiltonian is
modified by adding the term $P\Omega$, volume is represented
in terms of the strain tensor components as
\begin{eqnarray}\label{vol}
\Omega=\Omega_0 [1+\sum\limits_{\alpha =1}u_{\alpha\alpha}+
\sum\limits_{\alpha \neq
\beta}u_{\alpha\alpha}u_{\beta\beta}+O(u^3)]
\end{eqnarray}
and integration over homogeneous deformations is also
performed in Eq. (4). It was noted in [15] that the quadratic
terms in Eq. (5) can be important at high pressures
and for crystals with large striction effects. The
result is
\begin{eqnarray}\label{ogam}
&&H=\frac 12\int d^Dq(\tau _0+q^\sigma)\sum\limits_{a=1}^mS^a_qS^a_{-q}\\
&&+(u_0- \frac{z_0}{2})\sum\limits_{a=1}^m\int d^D\{q_i\}S^a_{q1}S^a_{q2}S^a_{q3}S^a_{-q1-q2-q3}\nonumber\\
&-&\frac{\delta}{2}\sum\limits_{a,b=1}^m\int
d^D\{q_i\}(S^a_{q1}S^a_{q2})(S^b_{q3}S^b_{-q1-q2-q3})\nonumber\\
&+&\frac{1}{2\Omega}(z_0 - w_0)\sum\limits_{a=1}^m\int
d^D\{q_i\}(S^a_{q1}S^a_{-q1})(S^a_{q2}S^a_{-q2}),\nonumber\\
&&z_0 =g_0 ^2/\lambda,  \ \  w_0 =g_0 ^{(0)2}/\lambda_0,  \nonumber
\end{eqnarray}
The effective coupling constant $v_0=u_0-z_0/2$, which
arises in the Hamiltonian owing to the striction effects
determined by $g_0$, can have positive and negative values.
Therefore, this Hamiltonian describes both first and
second-order phase transitions. When $v_0 = 0$, the
system exhibits tricritical behavior. Furthermore, the
effective interaction determined by the difference $z_0 –w_0$ in Eq. (6)
can also change the order of the phase transition.
This representation of the effective Hamiltonian
entails the existence of a higher order critical point
where tricritical curves intersect when $v_0 = 0$ and $z_0 =w_0$ [16].
Note that Hamiltonian (6) is isomorphic to the
Hamiltonian of the disordered Ising model with long-range
interactions under the tricritical condition $z_0 = w_0$.

Using the standard renormalization-group procedure
based on the Feynman diagram technique [17, 18]
with the propagator $G(\vec{k})=1/(\tau+|\vec{k}|^\sigma)$, we derive the
following expressions for the functions $\beta_v$,$\beta_\delta$,
$\beta_z$,$\beta_w$,$\gamma_\varphi$ and $\gamma_t$ specifying the
differential renormalization-group equation:
\begin{eqnarray}
    &&\beta_v=-(2\sigma-D)v\Big[1-36vJ_0+24\delta J_0\nonumber\\
    &&+1728\Big(2J_1-J_0^2-\frac29G\Big)v^2\nonumber\\
    &&-2304(2J_1-J_0^2-\frac 16G)v\delta\nonumber\\
    &&+672(2J_1-J_0^2-\frac23G)\delta ^2\Big],\nonumber\\
    &&\beta _\delta=-(2\sigma-D)\delta \Big[1-24vJ_0+16\delta J_0\nonumber\\
    &&+576(2J_1-J_0^2-\frac 23G_1)v^2-\nonumber\\
    &&-1152(2J_1-J_0^2-\frac13G)v\delta +352(2J_1-J_0^2-\frac 1{22}G)\delta ^2\Big],\nonumber
\end{eqnarray}
\begin{eqnarray}
    &&\beta _z=-(2\sigma-D)z \Big[1-24vJ_0-2zJ_0+8\delta J_0\nonumber\\
    &&+576(2J_1-J_0^2-\frac 23G)v^2-\nonumber\\
    &&-120(2J_1-J_0^2-\frac 85G)v\delta+96(2J_1-J_0^2-\frac 23G)\delta ^2\Big],\nonumber\\
    &&\beta _w=-(2\sigma-D)w \Big[1-24vJ_0+8\delta J_0-4zJ_0+2wJ_0\nonumber\\
    &&+576(2J_1-J_0^2-\frac 23G)v^2-\nonumber\\
    &&-120(2J_1-J_0^2-\frac 85G)v\delta
    +96(2J_1-J_0^2-\frac 23G)\delta ^2\Big].\nonumber\\
    &&\gamma_t=(2\sigma-D)\Big[-12vJ_0+4\delta J_0-2zJ_0+2wJ_0\nonumber\\
    &&+288\Big(2J_1-J_0^2-\frac13G\Big)v^2-\\
    &&-288(2J_1-J_0^2-\frac 23G)v\delta +32(2J_1-J_0^2-\frac12G)\delta ^2\Big], \nonumber\\
    &&\gamma_\varphi=(2\sigma-D)64G(3v^2-3v\delta+\delta^2),\nonumber\\
    &&J_1=\int \frac{d^Dq d^Dp}{(1+|\vec{q}|^\sigma)^2(1+|\vec{p}|^\sigma)
    (1+|q^2+p^2+2\vec{p}\vec{q}|^{\sigma/2})},\nonumber\\
    &&J_0=\int \frac{d^Dq}{(1+|\vec{q}|^\sigma)^2},\nonumber\\
    &&G=-\frac{\partial}{\partial |\vec{k}|^\sigma}\int \frac{d^Dq d^Dp}
    {(1+|q^2+k^2+2\vec{k}\vec{q}|^\sigma)(1+|\vec{p}|^\sigma)}\nonumber\\
    &&\cdot\frac1{(1+|q^2+p^2+2\vec{p}\vec{q}|^{\sigma/2})}\nonumber
\end{eqnarray}
In terms of the new effective interaction vertexes
\begin{equation}\label{vertex}
    v_1=v\cdot J_0,\ \ \ \ \  v_2=\delta\cdot J_0 \ \ \ \
    v_3=z\cdot J_0,\ \ \ \ \  v_4=w \cdot J_0.
\end{equation}
the functions $\beta_i$, $\gamma_\varphi$ and $\gamma_t$ are expressed as
\begin{eqnarray}\label{beta}
    &&\beta_1=-(2\sigma-D)\Big[1-36v_1+24v_2\\
    &&+1728\Big(2\widetilde{J_1}-1-\frac29\widetilde{G}\Big)v_1^2
    -2304(2\widetilde{J_1}-1-\frac 16\widetilde{G})v_1v_2\nonumber\\
    &&+672(2\widetilde{J_1}-1-\frac23\widetilde{G})v_2^2\Big],\nonumber\\
    &&\beta _2=-(2\sigma-D)\delta \Big[1-24v_1+8v_2\nonumber\\
    &&+576(2\widetilde{J_1}-1-\frac 23\widetilde{G})v_1^2
    -1152(2\widetilde{J_1}-1-\frac13\widetilde{G})v_1v_2\nonumber\\
    &&+352(2\widetilde{J_1}-1-\frac 1{22}\widetilde{G})v_2^2\Big],\nonumber\\
    &&\beta _3=-(2\sigma-D)v_3 \Big[1-24v_1+16v_2-2v_3\nonumber\\
    &&+576(2\widetilde{J_1}-1-\frac 23\widetilde{G})v_1^2
    -120(2\widetilde{J_1}-1-\frac 85G)v_1v_2\nonumber\\
    &&+96(2\widetilde{J_1}-1-\frac 23G)v_2^2\Big],\nonumber
    \end{eqnarray}
    \begin{eqnarray}
    &&\beta _4=-(2\sigma-D)v_4\Big[1-24v_1+8v_2-4v_3+2v_4\nonumber\\
    &&+576(2\widetilde{J_1}-1-\frac 23\widetilde{G})v_1^2
    -120(2\widetilde{J_1}-1-\frac 85G)v_1v_2\nonumber\\
    &&+96(2\widetilde{J_1}-1-\frac 23G)v_2^2\Big],\nonumber\\
    &&\gamma_t=(2\sigma-D)\Big[-12v_1+4v_2-2v_3+2v_4\nonumber\\
    &&+288\Big(2\widetilde{J_1}-1-\frac13\widetilde{G}\Big)v_1^2
    -192(2\widetilde{J_1}-1-\frac 23\widetilde{G})v_1v_2\nonumber\\
    &&+32(2\widetilde{J_1}-1-\frac12\widetilde{G})v_2^2\Big], \nonumber\\
    &&\gamma_\varphi=(2\sigma-D)64\widetilde{G}(3v_1^2-3v_1v_2+v_2^2).\nonumber
\end{eqnarray}
This redefinition is meaningful for $\sigma\leq D/2$. In this case,
$J_0$, $J_1$ and $G$ are divergent functions. Introducing the
cutoff parameter $\Lambda$, we obtain finite expressions for the
ratios $J_1/J_0^2$ and $G/J_0^2$ as $\Lambda\rightarrow\infty$.

The integrals are performed numerically. For $\sigma\leq D/2$,
a sequence of $J_1/J_0^2$ è $G/J_0^2$ corresponding to
various values of $\Lambda$ is calculated and extrapolated to
infinity.

It is well known that perturbation-theory series are
asymptotic and expressions (9) cannot be applied
directly since the interaction vertexes for fluctuations of
order parameters in the fluctuation region are too large.
For this reason, the required physical information was
extracted from these expressions by applying the Pade–Borel
method extended to the four-parameter case. The appropriate
direct and inverse Borel transforms have
the form
\begin{equation}
\begin{array}{rl} \displaystyle
  & f(v,\delta,z,w)=\sum\limits_{i_1,...,i_4}c_{i_1,...,i_4}v_1^{i_1}v_2^{i_2}v_3^{i_3}v_4^{i_4}\\
  &=\int\limits_{0}^{\infty}e^{-t}F(v_1t,v_2t,v_3t,v_4t)dt,\nonumber  \\
  & F(v,\delta,z,w)=\sum\limits_{i_1,...,i_4}\frac{\displaystyle c_{i_1,...,i_4}}{\displaystyle(i_1+...+i_4)!}v_1^{i_1}v_2^{i_2}v_3^{i_3}v_4^{i_4}.\nonumber
\end{array}
\end{equation}
To obtain an analytic continuation of the Borel transform
of a function, a series in an auxiliary variable $\theta$ is
introduced:
\begin{equation}
   {\tilde{F}}(v,\delta,z,w,\theta)=\sum\limits_{k=0}^{\infty}\theta^k\sum\limits_{i_1,...,i_4}\frac{\displaystyle c_{i_1,...,i_4}}{\displaystyle k!}v_1^{i_1}v_2^{i_2}v_3^{i_3}v_4^{i_4}\delta_{i_1+...+i_4,k}\  ,
\end{equation}
The Pade approximant [L/M] is applied to this series at
$\theta=1$. This procedure was proposed and tested in [19-22]
for describing the critical behavior of a number of systems
characterized by several interaction vertexes for
order-parameter fluctuations. It was found in [19-22]
that Pade approximants in the variable $\theta$ preserve the
symmetry of the system. This property is important for
analysis of multivertex models. The approximants [2/1]
are used to calculate the $\beta$ functions in the two-loop
approximation.

Critical behavior is completely determined by the
stable fixed points of the renormalization group transformation.
These points can be found from the condition
\begin{equation}\label{nep}
    \beta_i(v_1^*,v_2^*, v_3^*, v_4^*)=0 \ \ \ \ (i=1,2,3,4).
\end{equation}
The requirement of stability of a fixed point reduces to
the condition that the eigenvalues $b_i$ of the matrix
\begin{equation}  \displaystyle
B_{i,j}=\frac{\partial\beta_i(v_1^*,v_2^*, v_3^*, v_4^*)}{\partial{v_j}}.
\end{equation}
are positive.
\begin{table*}
\begin{center}
\begin{tabular}{|c|c|c|c|c|c|c|c|c|} \hline
$N$&$v_1^{*}$& $v_2^{*}$ & $v_3^{*}$& $v_4^{*}$ &$b_1$    & $b_2$   &$b_3$    & $b_4$ \\
\hline
\multicolumn{9}{|c|} {$\sigma=1.8$} \\
\hline
1&0.064189 &  0.046878&   0       &   0       &$0.626^*$& $0.626^*$& - 0.123 & - 0.123 \\
2&0.064189 &  0.046878&  0.066101 &   0       &$0.626^*$& $0.626^*$& 0.124  & 0.125 \\
3&0.064189 &  0.046878&  0.066101 &  0.066101 &$0.626^*$& $0.626^*$& 0.124  & - 0.124 \\
\hline
\multicolumn{9}{|c|} {$\sigma=1.9$} \\
\hline
4&0.066557 &  0.040818&    0      &   0       &$0.559^*$ & $0.559^*$ & - 0.118& - 0.118\\
5&0.066557 &  0.040818& 0.065716  &   0       &$0.559^*$ & $0.559^*$ & 0.119 & 0.119\\
6&0.066557 &  0.040818& 0.065716  & 0.065716  &$0.559^*$ & $0.559^*$ & 0.119 & - 0.119\\
\hline
\end{tabular} \end{center} \end{table*}

The exponent $\nu$ characterizing the growth of the correlation
radius near the critical point $(R_c\sim|T-T_c|^{-\nu})$ is
determined by the relation
\begin{eqnarray}
  \nu=\frac1\sigma(1+\gamma_t)^{-1}.\nonumber
\end{eqnarray}
The Fisher exponent $\eta$ describing the behavior of
the correlation function near the critical point in the
wave-vector space $(G\sim k^{2+\eta})$ is determined by the
scaling function $\gamma_\varphi$:
$ \eta=2-\sigma+\gamma_\varphi(v_1^*,v_2^*, v_3^*, v_4^*)$. The
remaining critical exponents can be determined from
the scaling relations.

The table shows the stable fixed points of the renormalization
group transformation and the fixed-point
eigenvalues of the stability matrix for $\sigma = 1.8$ and $1.9$. It
was shown in [13] that stable fixed points exist in the
physical region ($v_i^*>0$) for disordered systems only
when $\sigma\geq1.8$. When $\sigma<1.8$, the stable points of any
three-dimensional impurity system are characterized
by negative values of the vertex $v_1^*$.

An analysis of the critical points and their stability
shows that the fixed critical points of disordered systems
with long-range interactions (nos. 1 and 4) are
unstable under elastic deformations. The critical behavior
of disordered compressible systems with long-range
interactions is described by their respective fixed points
(nos. 2 and 5). The phase diagram of the substance can
contain tricritical points specified by fixed points
(nos. 3 and 6).

Calculations of the critical exponents for the stable
fixed points (nos. 2 and 5) yield
\begin{eqnarray}
   &&\sigma=1.9,\ \ \ \nu=0.721,\ \ \  \eta=0.134\\
   &&\sigma=1.8, \ \ \ \nu=0.758,\ \ \ \eta=0.251\nonumber
\end{eqnarray}
For the tricritical points (nos. 3 and 6), the critical exponents are
\begin{eqnarray}
   &&\sigma=1.9,\ \ \ \nu=0.686,\ \ \  \eta=0.134\\
   &&\sigma=1.8, \ \ \ \nu=0.721,\ \ \ \eta=0.251 \nonumber
\end{eqnarray}
Thus, elastic deformations change both critical and
tricritical behavior of three-dimensional disordered
Ising systems with long-range interactions.

The work is supported by Russian Foundation for Basic Research N
04-02-16002.

\newpage
\def\baselinestretch{1.0}

\end{document}